\documentstyle[eqsecnum,preprint,aps,prd]{revtex} 
\preprint{DAMTP-R96/21} 
\tighten
\begin{document}
\draft

\title{Relic Gravitational Waves from Cosmic Strings:
Updated Constraints and Opportunities for Detection}

\author{
R.R. Caldwell{$^1$}, 
R.A. Battye{$^{1,2}$}, 
and E.P.S. Shellard{$^1$}
}

\address{
\qquad\\
${}^1$
University of Cambridge, D.A.M.T.P.\\
Silver Street, Cambridge CB3 9EW, U.K. 
\qquad\\
\qquad\\
${}^2$
Theoretical Physics Group, Blackett Laboratory, Imperial College\\ 
Prince Consort Road, London SW7 2BZ, U.K.}
 
\maketitle
\begin{abstract} 

We examine the spectrum of gravitational radiation emitted by a network
of cosmic strings, with emphasis on the observational constraints and
the opportunities for detection. The analysis improves over past work,
as we use a phenomenological model for the radiation spectrum emitted
by a cosmic string loop. This model attempts to include the effect of
the gravitational back-reaction on the radiation emission by an
individual loop with a high frequency cut-off in the spectrum.
Comparison of the total spectrum due to a network of strings with the
recently improved bound on the amplitude of a stochastic gravitational
wave background, due to measurements of noise in pulsar signal arrival
times, allows us to exclude a range of values of $\mu$, the cosmic
string linear mass density, for certain values of cosmic string and
cosmological parameters.  We find the conservative bound $G\mu/c^2 <
5.4 (\pm 1.1) \times 10^{-6}$ which is consistent with all other
limits. We consider variations of the standard cosmological scenario,
finding that an under dense, $\Omega_0 < 1$ universe has little effect
on the spectrum, whereas the portion of the spectrum probed by
gravitational wave detectors is strongly sensitive to the thermal
history of the cosmological fluid. We discuss the opportunity for the
observation of this stochastic background by resonant mass and laser
interferometer gravitational wave detectors.

\end{abstract}


\section{Introduction}

Cosmic strings are line-like topological defects which may have formed
during a phase transition in the early universe
\cite{Historicref,Reviewref}.  Strings which formed with a
mass-per-unit-length $\mu$ such that $G\mu/c^2 \sim 10^{-6}$ may be
responsible for the formation of the large-scale structure and cosmic
microwave background anisotropy observed in the universe today. In
order to test the validity of the cosmic string scenario, it is
necessary to compare observations of our own universe with the
predictions of the cosmic string model. In this work we present updated
results of detailed computations of the spectrum of gravitational
radiation emitted by a network of cosmic strings
\cite{RadiationCalcref}.  We use these results to obtain limits on the
cosmic string mass-per-unit-length and to predict whether the
spectrum is within the sensitivity of forthcoming gravitational wave
detectors.

The spectrum of gravitational radiation due to cosmic strings has been
previously considered in detail in ref.\cite{RadiationCalcref} (and
references therein), where the methods used in the present paper were
described.  Here we benefit from recent work,
ref.\cite{BackReactionref}, which suggests that the effect of the
radiative back-reaction on a string loop is to damp out the higher
oscillation modes. Hence, the current work represents an improvement
due to the introduction of a phenomenological frequency cut-off, and
due to the updated observational constraints.

The most recent analysis of pulsar signal arrival times gives the limit
on the spectral density of gravitational radiation
\begin{equation}
\Omega_{\rm gr}(f) \equiv 
{f \over \rho_{\rm crit}}{d \rho_{\rm gr} \over d f} \mid_{f_{\rm obs}}\,\,
 <  9.3 \times 10^{-8} h^{-2}  \quad (95 \% \,{\rm CL}) \qquad \cite{MZVLref}
\label{GRBLimits}
\end{equation} 
in a logarithmic frequency interval at $f_{\rm obs} = (8 \, {\rm
yrs})^{-1}$.  This analysis corrects errors in previous  work
\cite{KTRref,TDref},  and uses an improved method for  testing the
hypothesis that the timing noise is due to a stochastic gravitational
wave background. For these reasons we use the latest bound, equation
(\ref{GRBLimits}), to obtain a constraint on $G\mu/c^2$ for given
values of the cosmic string and cosmological parameters.

The paper proceeds as follows. In section \ref{EmissionModelSection} we
present the model for the cosmic string loop radiation spectrum. This
model improves over past computations in that the effect of the
gravitational back-reaction is included. In section
\ref{AnalyticEstimateSection} we give an analytic estimate of the
spectrum of radiation emitted by a network of strings, allowing us to
study the model dependencies of the spectrum. In section
\ref{ObservationalBoundsSection} we present the results of our
numerical computation of the radiation spectrum.  Here we obtain limits
on $G\mu/c^2$ for given values of the cosmic string and cosmological
parameters.  In section \ref{DetectionSection} we discuss the
opportunities for the observation of the stochastic gravitational wave
background by the forthcoming generation of gravitational wave
detectors. We conclude in section \ref{ConclusionSection}.

\section{Cosmic String Loop Radiation Spectrum}
\label{EmissionModelSection}


The spectrum of gravitational radiation emitted by a network of cosmic
string loops is obtained using the product of a background $\Omega = 1$
FRW cosmological model, an extended one-scale model for the evolution
of a network of cosmic strings, and a model of the emission of
gravitational radiation by cosmic string loops. The procedure by which
the spectrum is computed has been presented in detail in
ref.\cite{RadiationCalcref}. In this section we discuss the model for
radiation by an individual loop.

The model of the emission of gravitational radiation by cosmic string
loops is composed of the following three elements.
\begin{enumerate}
\item 
A loop radiates with power $P = \Gamma G \mu^2 c$.  The dimensionless
radiation efficiency, $\Gamma$, depends only on the loop configuration,
rather than overall size. Recent studies of realistic loops indicate
that the distribution of values of the efficiency has a mean value 
$\langle\Gamma\rangle \approx 60$ \cite{CasperCommref}.
\item 
The frequency of radiation emitted by a loop of invariant length $L$ is
$f_n = 2 n/L$ where $n=1,2,3,...$ labels the oscillation mode.  \item
The fraction of the total power emitted in each mode of oscillation $n$
at frequency $f_n$ is given by the coefficient $P_n$ where
\begin{equation}
P =  \Big(\sum_{n=1}^\infty P_n \Big) G \mu^2 c = \Gamma G \mu^2 c.
\label{EmissionModel}
\end{equation}
Analytic and numerical studies suggest that the radiation efficiency
coefficients behave as $P_n \propto n^{-q}$ where $q$ is the spectral
index.
\end{enumerate}
This model has several shortcomings. First, the spectral index $q$ has
not been well determined by the numerical simulations. Numerical work
suggests $q = 4/3$ \cite{CuspyIndexref} as occurs with cuspy loops,
loops along which points momentarily reach the velocity of light, based
on simulations of a network of cosmic strings. However, these
simulations have limited resolution of the important small scale
features of the long strings and loops.  Hence, the evidence for
$q=4/3$ is not compelling. Analytic work suggests that $q = 2$
\cite{KinkyIndexref}, characteristic of kinky loops, loops along which
the tangent vector changes discontinuously as a result of
intercommutation, may be more realistic. Second, the effect of
back-reaction on the motion of the string has been ignored. In this
model, a loop radiates at all times with a fixed efficiency, in all
modes, until the loop vanishes.  As we shall next argue, the
back-reaction will result in an effective high frequency cut-off in the
oscillation mode number.  Thus, the loop will only radiate in a finite
number of modes, and hence in a finite range of frequencies. The
resolution of these issues may have strong consequences for the entire
spectrum produced by a network of strings.

Recent advances in the understanding of radiation back-reaction on
global strings suggest various modifications to the simplified model of
emission by cosmic string loops. There are remarkable similarities
between gravitational radiation and Goldstone boson radiation from
strings \cite{DetailedRadiationCalcref}, which we believe allow us to
make strong inferences as to the nature of gravitational radiation
back-reaction.  For example, the same, simple model for the emission of
gravitational radiation by cosmic strings may be transferred over to
global strings:  in the absence of Goldstone back-reaction,  global
string loops radiate at a constant rate, at wavelengths given by even
sub-multiples of the loop length, with an efficiency as described by an
equation similar to (\ref{EmissionModel}).  Hence, our argument
proceeds as follows. Fully relativistic field theory simulations of
global strings have been carried out \cite{BackReactionref}, where it
was observed that the power in high oscillation modes is damped by the
Goldstone back-reaction on periodic global strings.  An analytic model
of Goldstone back-reaction \cite{BackReactionref}, as a modification of
the classical Nambu-Goto equations of motion for string, was developed
which successfully reproduces the behaviour observed in field theory
simulations. That is, high frequency modes are damped rapidly, whereas
low frequency modes are not.  Thus, we are motivated to rewrite
equation (\ref{EmissionModel}) for global strings, and by analogy for
cosmic strings, as \begin{equation} P=\Big( \sum_{n=1}^{n_*}P_n \Big) G
\mu^2 c =\Gamma G\mu^2 \end{equation} where $n_*$ is a cut-off
introduced to incorporate the effects of back-reaction. By comparing
the back-reaction length-scale to the loop size, we estimate that such
a cut-off should be no larger than $\sim (\Gamma G \mu/c^2)^{-1}$.  The
ongoing investigations of global and cosmic string back-reaction
\cite{BackReactionResearchref} have not yet reached the level of
precision where a firm value of $n_*$ may be given. As we demonstrate
later, the effect on the radiation spectrum is significant only for
certain values of the cut-off.

\section{Analytic Estimate of the Radiation Spectrum}
\label{AnalyticEstimateSection}

Analytic expressions for the spectrum of gravitational radiation
emitted by a network of cosmic strings have been derived in
ref.\cite{RadiationCalcref}. While these analytic expressions are
simplified for convenience, they offer the opportunity to examine the
various dependencies of the spectrum on cosmic string and cosmological
parameters.

The spectrum of gravitational radiation produced by a network of cosmic
strings has two main features. First is the `red noise' portion of the
spectrum with nearly equal gravitational radiation energy density per
logarithmic frequency interval, spanning the frequency range $10^{-8}\,
{\rm Hz} \lesssim f \lesssim 10^{10}\,{\rm Hz}$. This spectrum corresponds
to gravitational waves emitted during the radiation-dominated expansion
era. This feature of the spectrum may be accessible to the forthcoming
generation of gravitational wave detectors. Second is the peak in the
spectrum near $f \sim 10^{-12} \, {\rm Hz}$. The amplitude and slope of
the spectrum from the peak down to the flat portion of the spectrum is
tightly constrained by the observed limits on pulsar timing noise.

\subsection{Red Noise Portion of the Spectrum}

An analytic expression for the `red noise' portion of the gravitational
wave spectrum is given as follows:
\begin{equation}
{f \over \rho_{\rm crit}}{d \rho_{\rm gr} \over d f}
= {8 \pi \over 9} A {\Gamma(G\mu)^2 \over \alpha c^4}
\bigl[1 - {\langle v^2\rangle / c^2}\bigr] {(\beta^{-3/2} - 1) \over (z_{\rm eq} + 1)}
\qquad 10^{-8}\, {\rm Hz} \lesssim f \lesssim 10^{10}\,{\rm Hz} 
\label{RedNoiseSpectrum}
\end{equation}
\begin{equation}
A  \equiv   \rho_{\infty} d_H^2(t) c^2 / \mu \qquad\qquad
\beta  \equiv   \bigl[1 + f_{\rm r}\,\alpha d_H(t) c/(\Gamma G \mu t)\bigr]^{-1}
\end{equation}
In the above expressions, $\rho_{\infty}$ is the energy density in
`infinite' or long cosmic strings, $\alpha$ is the invariant length of
a loop as a fraction of the physical horizon length $d_H(t)$ at the
time of formation, $\langle v^2 \rangle$ is the rms velocity of the
long strings, and $f_{\rm r}\approx 0.7$ is a correction for the
damping of the relativistic center-of-mass velocity of newly formed
string loops. All quantities are evaluated in the radiation era;
$d_H(t) = 2 c t$, $A = 52 \pm 10$, and $\langle v^2 \rangle/c^2 = 0.43
\pm 0.02$ \cite{Simulationref}.

The above expression for the spectrum has been obtained assuming no
change in the number of relativistic degrees of freedom, $g$, of the
background radiation-dominated fluid. However, the annihilation of
massive particle species as the cosmological fluid cools leads to a
decrease in the number of degrees of freedom, and a redshifting of all
relativistic particles not thermally coupled to the fluid. This has the
effect of modifying the amplitude of the spectral density
\cite{BennettPapersref}, equation (\ref{RedNoiseSpectrum}), by a factor
$(g(T_f)/g(T_i))^{1/3}$ where $g(T_{i,f})$ is the number of degrees of
freedom at temperatures before and after the annihilations.  Using a
minimal GUT particle physics model as the basis of the standard thermal
history, we see that the red noise spectrum steps downwards with
growing frequency.
\begin{eqnarray}
{f \over \rho_{\rm crit}}{d \rho_{\rm gr} \over d f}
&=& {8 \pi \over 9} A {\Gamma(G\mu)^2 \over \alpha c^4}
\bigl[1 - \langle v^2\rangle/c^2\bigr] {(\beta^{-3/2} - 1) \over (z_{\rm eq} + 1)}
\cr\cr
&&\times\cases{
1	&  \cr
\qquad  
10^{-8}\, {\rm Hz} \lesssim f \lesssim 10^{-10}\alpha^{-1}
	\, {\rm Hz} & \cr\cr
(3.36/10.75)^{1/3} = 0.68 & \cr	
\qquad  
10^{-10}\alpha^{-1}\, {\rm Hz} \lesssim f \lesssim 10^{-4}\alpha^{-1}
	\, {\rm Hz} & \cr\cr
(3.36/106.75)^{1/3}= 0.32 & \cr
\qquad
10^{-4}\alpha^{-1}\, {\rm Hz} \lesssim f \lesssim 10^{8}
	\, {\rm Hz} & }
\label{RedNoiseSpectrumToday}
\end{eqnarray}
Hence, the red noise spectrum is sensitive to the thermal history of
the cosmological fluid. The locations of the steps in the spectrum are
determined by the number of relativistic degrees of freedom as a
function of temperature, $g(T)$. As an example, we present the effect
of a non-standard thermal history on the spectrum in Figure
\ref{figure1}. In this sample model, the number of degrees of freedom
$g(T)$ decreases by a factor of $10$ at the temperatures $T=10^9,\,
10^5,\, 1\, {\rm GeV}$. The effect on the spectrum is a series of steps down
in amplitude with increasing frequency; detection of such a shift would
provide unique insight into the particle physics content of the early
universe at temperatures much higher than may be achieved by
terrestrial particle accelerators. In the case of a cosmological model
with a thermal history such that $g(T_i) \gg g(T_f)$ for $T_i > T_f$,
all radiation emitted before the cosmological fluid cools to $T_i$ will
be redshifted away by the time the fluid reaches $T_f$. As we discuss
later, such a sensitivity of the spectrum to the thermal history
affects the nucleosynthesis bound on the total energy in  gravitational
radiation, and the opportunity to detect high frequency gravitational
waves.

\vskip 0.5in
\subsection{Peaked Portion of the Spectrum}

We now turn our attention to the peaked portion of the gravitational
wave spectrum. The shape of this portion depends on the model for the
emission by a loop, presented in section \ref{EmissionModelSection}.
The dominant behaviour of the peaked portion of the spectrum is given by
\begin{eqnarray}
{f \over \rho_{\rm crit}}{d \rho_{\rm gr} \over d f}
 &\approx & \cases{
C_1 /f^{(q-1)}	&	$1 < q < 2$ \cr
C_2 /f		&	$q \ge 2$ } \cr\cr
&& {\rm for}\quad 10^{-12} \, {\rm Hz} \lesssim f \lesssim 10^{-8} \, {\rm Hz}.
\label{Peaked Spectrum}
\end{eqnarray}
Here $C_{1,2}$ are dimensionful quantities which depend on $G\mu/c^2,
\, \alpha, \,\Gamma, \, A, \, q$ and $n_{*}$. A lengthy expression
displaying the full dependence of the spectrum on these parameters is
not particularly enlightening.  However, the qualitative behaviour,
described in more detail  in ref.\cite{DetailedRadiationCalcref}, is as
follows.  The overall height of the spectrum depends linearly on
$G\mu/c^2$, while the frequency at which the peaked spectrum gives way
to the red noise spectrum depends inversely on $\alpha$.  The important
result is that for values of the mode cut-off $n_{*} \lesssim 10^{2}$,
the spectrum drops off as $1/f$ for any value $q \ge 4/3$. As a
demonstration, sample spectra with various values of $n_{*}$ are
displayed in Figure \ref{figure2}. Hence, the introduction of a
sufficiently low mode cut-off eliminates the dependence of the spectrum
on $q$, the loop spectral index.

We have also examined the spectrum of gravitational radiation produced
by the cosmic string network in an open FRW space-time, with $0.1 <
\Omega_0 < 1$. For this range of values of the cosmological density
parameter, the portion of the spectrum produced at a time $t$ is
shifted downward by a factor $\Omega(t)$. Sample spectra for various
values of $\Omega_0$ are displayed in Figures
\ref{figure3a}-\ref{figure3b}.  In the case that the spectrum drops off
slower than $1/f$, the spectral density at frequencies as high as $f
\sim 10^{-5}\,{\rm Hz}$ is diluted for $\Omega_0 < 1$. In the case that
the spectrum drops off as $1/f$,  only at lower frequencies, $f
\lesssim 10^{-10}\,{\rm Hz}$, is the spectral density affected.

\section{Observational Bounds on the Radiation Spectrum}
\label{ObservationalBoundsSection}

In this section we determine the observational constraint on the cosmic
string mass-per-unit-length $G\mu/c^2$. To begin, we discuss the recent
analyses of the pulsar timing data, after which we apply the newly
obtained bounds to the cosmic string gravitational wave background.

The observations used to place a limit on the amplitude of a stochastic
gravitational wave background consist of pulse arrival times for PSR
B1937+21 and PSR B1855+09 \cite{KTRref}. Although there has been some
recent  controversy regarding the analysis of the pulsar timing data
\cite{TDref}, the work by McHugh {\it et al} \cite{MZVLref} best
assesses the likelihood that the timing residuals are due to
gravitational radiation. We note that all analyses to date have assumed
a flat, red noise  spectrum for the gravitational wave spectral
density. Such an assumption  is only justified for a restricted range
of frequencies in the case of  a background due to cosmic strings, as
we have demonstrated in the  preceding section. Hence, a statistical
analysis which uses a realistic  model of the cosmic string spectrum
may obtain a different limit on the  amplitude of the spectral
density.

We now present values of the parameter $G\mu/c^2$ for values of
$\alpha$ which satisfy the pulsar timing constraint on the
gravitational radiation spectrum. Contours of constant $\Omega_{\rm
gr}$ in the logarithmic frequency bin $f = (8\, {\rm yrs})^{-1}$, given
by (\ref{GRBLimits}), in $(\alpha,G\mu/c^2)$ parameter space, are shown
in Figure \ref{figure4}. We have used cosmological parameters $\Omega_0
= 1$ and $h \in [0.5,\,0.75]$ with the cosmic string loop radiation
efficiency $ \Gamma  = 60$.   We find
\begin{equation}
G\mu/c^2 < \cases{ 
2.0 (\pm 0.4) \times 10^{-6} \,\, (2 h)^{-8/3}	& $q = 4/3$ \cr 
5.4 (\pm 1.1) \times 10^{-6}	& $q \ge 2$ or $n_{*} \lesssim 10^{2}$.} 
\label{pulsarbound}
\end{equation}
These constraints correspond to the maximum value of $G\mu/c^2$ along
the contour of constant  $\Omega_{\rm gr}$.  In the case $q=4/3$, this
maximum occurs near $\alpha =  \Gamma  G \mu/c^2$, the expected size of
newly formed loops based on considerations of the gravitational
back-reaction, while for the $q \ge 2$ or $n_{*} \lesssim 10^{2}$ case,
the maximum occurs at a slightly smaller value of $\alpha$.  For both
larger and smaller values of $\alpha$ the bounds become more stringent,
as described in \cite{RadiationCalcref}.  The unusual dependence on $h$
is due to the contribution from high mode number waves emitted in the
matter era, for which the amplitude depends on both the slope of the
spectrum and the time of radiation-matter equality.  The quoted errors
are due to uncertainties in the cosmic string model parameters measured
by the numerical simulations \cite{Simulationref}.  For
the case of an open universe, there is no change in the $q \ge 2$ or
$n_{*} \lesssim 10^{2}$ bound. However, the $q=4/3$ bound is weakened
by a factor $\sim 1/\Omega_0$.

We now comment on the validity of the model which we have used to
generate the gravitational radiation spectra.  We have shown that the
observational bounds on the total spectrum are sensitive to the value
of the loop spectral index $q$, unless there is a back-reaction cut-off
$n_{*} \lesssim 10^2$.  Furthermore, we have noted that there is
uncertainty in the characteristic value of the loop spectral index,
$q$.  Hence, we feel that it is more reasonable to take the
conservative bound of (\ref{pulsarbound}) at the present.  Next,
consider the extended one-scale model, described in
refs.\cite{Reviewref,RadiationCalcref}, for the evolution of the string
network. This model assumes that the long string energy density scales
relative to the background energy density, with the dominant energy
loss mechanism due to the formation of loops of a characteristic
scale.  A more sophisticated model, by Austin {\it et al}
\cite{ACKModelref}, attempts to include the effect of the gravitational
back-reaction on the long-term evolution of the string network; results
suggest that an effect of the back-reaction may be to lower the scaling
density in long strings at late times, beyond the reach of numerical
simulations. Hence, there is some uncertainty as to how accurately the
extended one-scale model describes the evolution of the string
network.  However, we do not believe that these considerations could
result in a decrease in the amplitude of the gravitational wave
background by more than $\sim 50\%$.  Thus, we quote $G\mu /c^2 < 5.4
(\pm 1.1) \times 10^{-6}$ as a conservative bound on the cosmic string
mass-per-unit-length.

The bounds computed in ref. \cite{RadiationCalcref} due to the
constraint on the total energy density in gravitational waves at the
time of nucleosynthesis remain valid. For a limit on the effective
number of neutrino species $N_\nu < 3.1,\, 3.3,\, 3.6$, the bound on
the cosmic string mass-per-unit-length is $G\mu/c^2 < 2,\, 6,\, 10
\times 10^{-6}$ respectively, evaluated at $\alpha = \Gamma G\mu/c^2$.
The big-bang nucleosynthesis limit on the number of effective neutrino
species is a conservative $N_\nu < 4$, owing to uncertainties in the
systematic errors in the observations of light element abundances
\cite{BBNLimitsref}.  Hence, until the observations are refined, the
nucleosynthesis bound is weaker than the pulsar timing bound.
Furthermore, the translation of the limit on $N_\nu$ into the bound on
$G\mu/c^2$ is sensitive to the thermal history of the cosmological
fluid \cite{BennettPapersref}. The bound on the string
mass-per-unit-length may be considerably weakened if the cosmological
fluid possessed many more relativistic degrees of freedom in the early
universe beyond those given by a minimal GUT model.

Comparing detailed computations of the large angular scale cosmic
microwave background temperature anisotropies induced by cosmic strings
\cite{CMBref} with observations, the cosmic string mass-per-unit-length
has been normalized to \begin{equation} G\mu/c^2 =
1.05^{\,+0.35}_{\,-0.20}\, \times 10^{-6}.
\label{NormalizationEquation} \end{equation} Therefore, given the
uncertainties in the extended one-scale model, we find the
gravitational radiation spectrum to be compatible with observations.

\section{Detection of the Radiation Spectrum} \label{DetectionSection}

We would like to determine whether the stochastic gravitational wave
spectrum emitted by cosmic strings may be observed by current and
planned detectors.  Because all ground-based detectors operate at
frequencies $f \gtrsim 10^{-3}\, {\rm Hz}$, we need only consider the
`red noise' portion of the gravitational wave spectrum
(\ref{RedNoiseSpectrumToday}). Noting that the spectral density,
$\Omega_{\rm gr}(f)$, has a minimum value when $\alpha \to 0$ (this has
been pointed out by Allen in \cite{LesHouchesref}) the predicted
spectrum is bounded from below by
\begin{eqnarray}
\Omega_{\rm gr}(f) &\ge &  {24 \pi \over 9} A f_{\rm r} 
{(1 - \langle v^2 \rangle/c^2 ) \over (1 + z_{\rm eq})} (G\mu/c^2)
\Bigl({g(T_0) / g(T_{\rm GUT})}\Bigr)^{1/3} \cr\cr
&\ge & 1.4 \times 10^{-9} \qquad
{\rm for}\quad 10^{-8}\, {\rm Hz} \lesssim f \lesssim 10^{10}\, {\rm Hz}.
\label{MinimumSpectrum}
\end{eqnarray}
Here we have used the normalization in (\ref{NormalizationEquation})
for $G\mu/c^2$, Hubble parameter $h = 0.75$, and assumed a minimal GUT
thermal history.  Hence, this lower bound is valid up to frequencies $f
\sim 10^{-3} \alpha^{-1}\,{\rm Hz} \sim 10 \, {\rm Hz}$ based on our
knowledge of the number of relativistic degrees of freedom, $g$,
of the primordial fluid up to temperatures $T \sim 10^3 \, {\rm GeV}$.
Notice that a measurement of the spectral density due to cosmic strings
at higher frequencies would sample $g$ at higher temperatures.

We may in turn place a lower bound on
the amplitude of the dimensionless strain predicted for the
gravitational wave emitted by cosmic strings:
\begin{eqnarray}
h_{\rm c} &=& 1.3 \times 10^{-18} h \sqrt{\Omega_{\rm gr}(f)} 
\Bigl({f \over 1 \, {\rm Hz}}\Bigr)^{-1} \cr\cr
&\ge &3.6 \times 10^{-23}  
\Bigl({f \over 1 \, {\rm Hz}}\Bigr)^{-1}
\qquad
{\rm for}\quad 10^{-6}\, {\rm Hz} \lesssim f \lesssim 10^{8}\, {\rm Hz}.
\label{MinimumStringStrain}
\end{eqnarray} 
The expressions (\ref{MinimumSpectrum}-\ref{MinimumStringStrain}) are
useful for comparison with the planned sensitivities of the forthcoming
generation of gravitational wave detectors \cite{AmaldiConfref}.

The most promising opportunity to probe for a stochastic gravitational
wave background due to cosmic strings is through a cross-correlation of
the observations of the advanced LIGO, VIRGO and LISA interferometers.
It is estimated that the advanced LIGO detectors will have the sensitivity
\begin{equation}
h_{\rm 3/yr} = 5.2 \times 10^{-25} \Bigl({f \over 1 \, {\rm k Hz}}\Bigr)^{1/2}
\end{equation}
for stochastic waves (equation 125c of \cite{LIGOref}), sufficient to
measure the minimum predicted strain (\ref{MinimumStringStrain}) near
$f \sim 100 \,{\rm Hz}$ in a $1/3$-year integration time.  More recent
calculations \cite{LIGOPosref} confirm that the orientation of the advanced LIGO
interferometers will be sufficient in order to detect the cosmic string gravitational
wave background. For the LISA project \cite{LISAref}, comparison of the
projected strain sensitivity $h_c \sim 10^{-20}$ at the frequency  $f
\sim 10^{-3}\, {\rm Hz}$ \cite{LISAref} with (\ref{MinimumStringStrain})
indicates that the space-based interferometer will be capable of
detecting a gravitational radiation background produced by a network of
cosmic strings.

Other ground-based interferometric gravitational wave detectors are in
development or under construction. The GEO600 and TAMA300 detectors,
operating near frequencies $f \sim 10^3 \, {\rm Hz}$, may also be
capable of measuring a cosmic string generated background.

A network of resonant mass antennae, such as bar and TIGA detectors may
probe for a stochastic background.  Successful detection by these
antennae will require improved sensitivity and longer integration time.
However, cross-correlation between a narrow-band bar and a wide-band
interferometric detector may improve the opportunities.  Estimates of
the sensitivity of such a system, assuming optimum detector alignment
\cite{Correlationref}, indicate that 
\begin{equation}
\sqrt{h_{\rm int} h_{\rm b}} \gtrsim 
2.6 \times 10^{-19} \sqrt{\Omega_{\rm gr}(f)}
\Bigl({f \over 1 \, {\rm k Hz}}\Bigr)^{-3/2} 
\Bigl({t_{\rm obs} \over 10^7 \, {\rm s}}\Bigr)^{1/2} 
\end{equation}
is necessary to detect a background $\Omega_{\rm gr}$.  Hence, for a
$1/3$-year observation time, the bar and interferometer strain
sensitivities at $1 \, {\rm kHz}$ must be better than $\sim 10^{-23}$
in order to detect the cosmic string background.

We stress that the amplitude of the cosmic string gravitational wave
background for frequencies  $f \gtrsim 10\,{\rm Hz}$ is sensitive to the
number of degrees of freedom of the cosmological fluid at temperatures
$T \gtrsim 10^3 \,{\rm GeV}$. The amplitude of the cosmic string
background at LISA-frequencies, near $f \sim 10^{-3}\, {\rm Hz}$, is
firm, since the cosmological fluid near the temperature $T \sim 10
\,{\rm MeV}$ is well understood.  However, at the higher frequencies
probed by ground-based detectors, our uncertainty in the number of
degrees of freedom of the cosmological fluid, as determined by the
correct model of particle physics at that energy scale, may reduce the
predicted amplitude (\ref{MinimumSpectrum}) of gravitational
radiation.

\section{Conclusion}
\label{ConclusionSection}

In this paper, we presented improved calculations of the spectrum of
relic gravitational waves emitted by cosmic strings.  We demonstrated
that the effect of a gravitational back-reaction on the radiation
spectrum of cosmic string loops, for which there is an effective mode
cut-off $n_{*} \lesssim 10^2$, may serve to weaken the pulsar timing
bound on the cosmic string mass-per-unit-length.  Arguing for a model
of radiation by loops, for which either the spectral index is $q \ge 2$
or there is an emission mode cut-off $n_{*} \lesssim 10^2$, we obtain
the conservative bound $G \mu/c^2 < 5.4 (\pm 1.1) \times 10^{-6}$ due
to observations of pulsar timing residuals. We believe this bound to be
robust, as the spectrum depends weakly on the precise value of the mode
cut-off, up to $n_{*} \sim 10^2$.  We have noted the interesting result
that the flat, red noise portion of the gravitational wave spectrum is
sensitive to the thermal history of the cosmological fluid, revealing
features of the particle physics content at early times. Finally, we
have pointed out that the generation of advanced LIGO, VIRGO and LISA
interferometers should be capable of detecting the predicted stochastic
gravitational wave background due to cosmic strings.

\acknowledgements

The work of RRC and EPSS is supported by PPARC grant GR/K29272. The work
of RAB is supported by PPARC postdoctoral fellowship grant GR/K94799.

\begin{figure}
\caption{
The effect of a non-standard thermal history of the cosmological fluid
on the amplitude of the red noise portion of the gravitational wave
spectrum is shown. The solid curve displays the spectrum produced using
a minimal GUT with a maximum $g = 106.75$.  The dashed curve shows the
spectrum produced allowing for a hypothetical, non-standard evolution
of $g(T)$, as might occur if there were a series of phase transitions,
or a number of massive particle annihilations as the universe cooled.
For temperatures $T > 10^9 \,{\rm GeV}$, the number of degrees of
freedom is $g = 10^4$.  For $10^5 \, {\rm GeV} < T < 10^9 \, {\rm
GeV}$, $g = 10^3$.  For $T<10^5 \, {\rm GeV}$, the standard thermal
scenario is resumed. }
\label{figure1}
\end{figure}

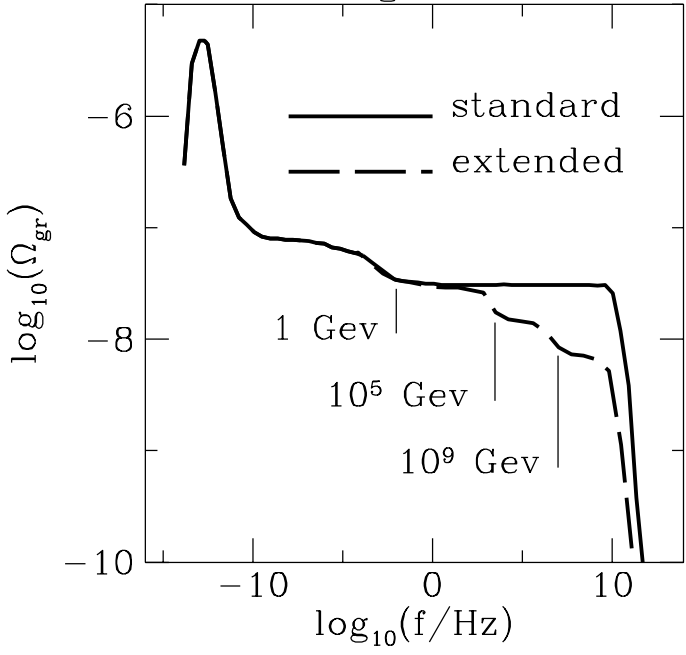
\begin{figure}
\caption{
The effect of a cut-off in the radiation mode number on the spectrum of
gravitational radiation is shown. Curves for the loop radiation
spectral index $q=2,\, 4/3$ for various values of $n_{*}$ are shown.
The vertical line shows the location of the frequency bin probed by
pulsar timing measurements. For $n_{*} {\protect{\lesssim}} 10^2$ the
shape of the spectrum is insensitive to the value of $q$ for purposes
of pulsar timing measurements. For increasing $n_{*}$, more radiation
due to late-time cosmic string loops is emitted in the pulsar timing
frequency band. }
\label{figure2}
\end{figure}

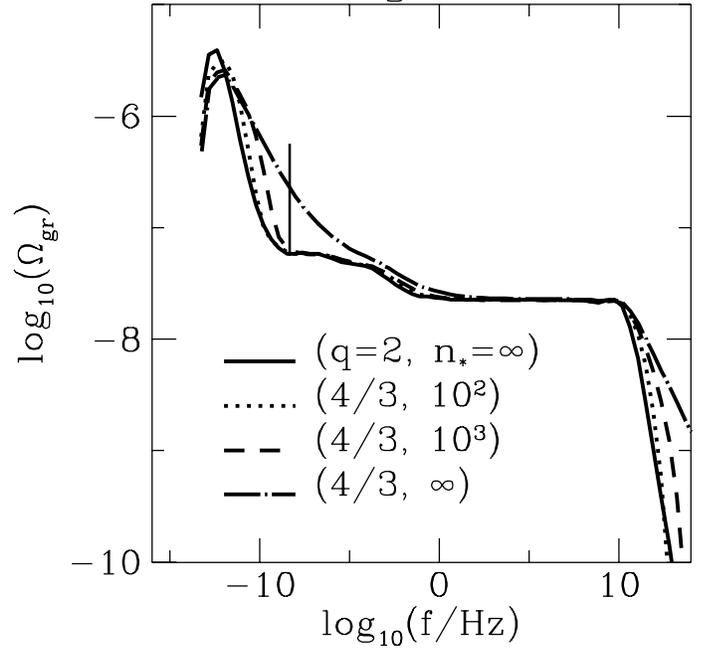
\begin{figure} 
\caption{ 
The effect of a low density, $\Omega_0 < 1$ universe on the peaked
portion of the gravitational wave spectrum.  The solid, long- and
short-dashed curves represent spectra for $\Omega_0 = 1, \,0.6,
\,0.2$.  The vertical line shows the location of the frequency bin
probed by pulsar timing measurements. For the loop spectral index
$q=2$, a low density universe dilutes only the lowest frequency waves,
corresponding the radiation emitted by loops still present today.}
\label{figure3a} 
\end{figure}

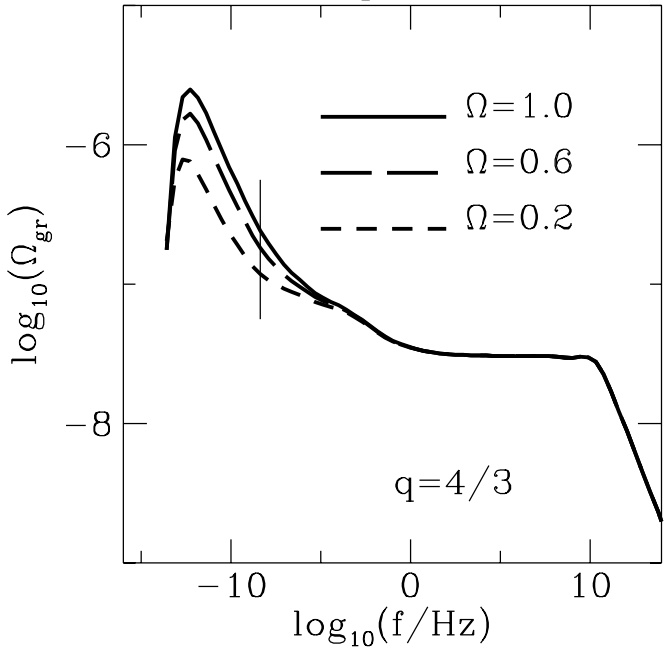
\begin{figure} 
\caption{ 
The effect of a low density, $\Omega_0 < 1$ universe on the peaked
portion of the gravitational wave spectrum.  The solid, long- and
short-dashed curves represent spectra for $\Omega_0 = 1, \,0.6,
\,0.2$.  The vertical line shows the location of the frequency bin
probed by pulsar timing measurements. For the loop spectral index
$q=4/3$, a low density universe leads to a dilution of gravitational
waves with wavelengths up to $f \sim 10^{-5}\, {\rm Hz}$. }
\label{figure3b} 
\end{figure}

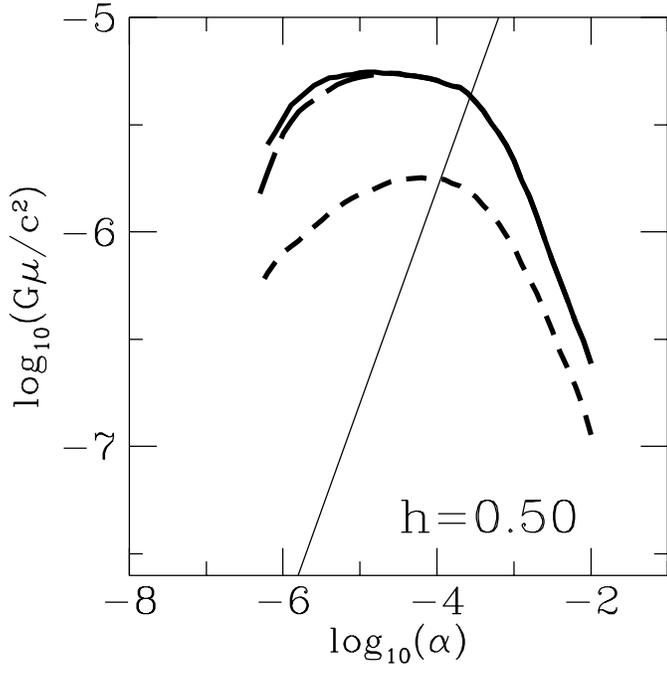
\begin{figure}
\caption{
Curves of constant $\Omega_{\rm gr}$ in $(\alpha, G\mu/c^2)$ parameter
space are shown. For a given value of $\alpha$, these figures give the
observational bound on $G\mu/c^2$ in the case $h=0.5, \, 0.75$. In each
figure, the constraining curves for $q=10,\, 2,\, 4/3$ are given by the
solid, long-, and short-dashed curves. The light dashed lines show
$\alpha = \Gamma G\mu/c^2$. The most conservative constraint is
$G\mu/c^2 < 5.4 \times 10^{-6}$.  }
\label{figure4}
\end{figure}


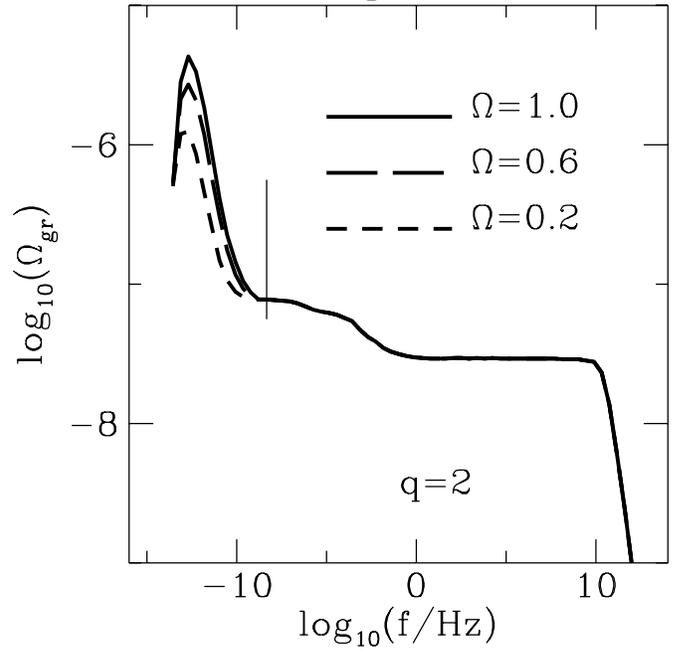
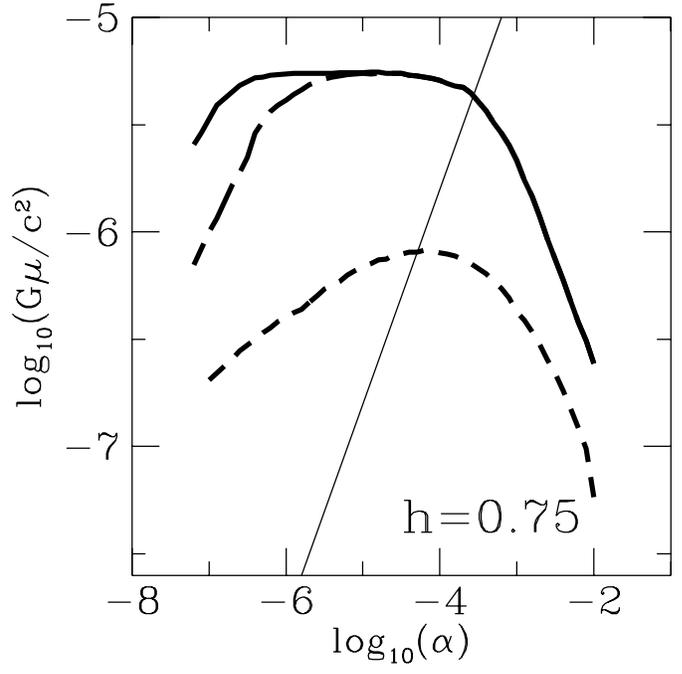
\end{document}